\documentclass{elsart}

\usepackage{graphicx,amsfonts,amsmath}

\begin{document}
\begin{frontmatter}

\title{\bf Particle In Cell Simulation of Combustion Synthesis of TiC Nanoparticles}
\author{G. Zuccaro$^{1}$, G. Lapenta$^{2,3}$, G. Maizza$^{1,2}$} \address{ $^1$Dipartimento di Scienza dei Materiali e
Ingegneria Chimica \\ Politecnico di Torino, Italy\\ $^2$Istituto
Nazionale per la Fisica della Materia (INFM), Sezione di Torino,
Italy \\ $^3$Plasma Theory Group, Theoretical Division\\
Los Alamos National Laboratory, University of California,  USA. }
\ead{lapenta@lanl.gov}

\begin{abstract}
A coupled continuum-discrete numerical model is presented to study
the synthesis of TiC nanosized aggregates during a
self-propagating combustion synthesis (SHS) process. The overall
model describes the transient of the basic mechanisms governing
the SHS process in a two-dimensional micrometer size geometry
system. At each time step, the continuum (micrometer scale) model
computes the current temperature field according to the prescribed
boundary conditions. The overall system domain is discretized with
a desired number of uniform computational cells. Each cell
contains a convenient number of computation particles which
represent the actual particles mixture. The particle-in-cell
(discrete) model maps the temperature field from the (continuum)
cells to the respective internal particles. Depending on the
temperature reached by the cell, the titanium particles may
undergo a solid-liquid transformation. If the distance between the
carbon particle and the liquid titanium particles is within a
certain tolerance they will react and a TiC particle will be
formed in the cell. Accordingly, the molecular dynamic method will
update the location of all particles in the cell and the amount of
transformation heat accounted by the cell will be entered into the
source term of the (continuum) heat conduction equation. The new
temperature distribution will progress depending on the cells
which will time-by-time undergo the chemical reaction. As a
demonstration of the effectiveness of the overall model some
paradigmatic examples are shown.

\end{abstract}

\end{frontmatter}

\section{Introduction}

The self-propagating high-temperature synthesis process (SHS)
\cite{Merzhanov_2} is a promising method employed for the
synthesis of many advanced materials,
 such as ceramic, intermetallics, composites, etc. \cite{Suryanarayama}.
Usually the process applies to powders mixture which are
conveniently homogenized and pressed  in order to form a loosely
compacted pellet. The SHS exploits the ability of certain
materials mixture in producing high exothermic and self-sustaining
reactions once ignited locally or uniformly. Ignition can be made
by a laser beam, induction, resistance, radiant or a spark source.
The exothermic reaction makes the temperature increase rapidly in
the pellet, reaching and surpassing the combustion temperature.
The resulting product are usually very pure and rather porous
(about 50\% of the theoretical density \cite{Suryanarayama}). The
final product composition and its morphology depend on
\cite{Weiner}:

\begin{description}
    \item[i)] initial
particles size and distribution, shape and purity;
    \item[ii)] initial
density of the reacting mixture;
    \item[iii)] initial temperature of the
reacting system;
    \item[iv)] size of the sample and reactor configuration;
    \item[v)] dilution of the reacting mixture with the final
    product.
\end{description}Thus, the actual combustion process involves simultaneously
critical factors at both the particle (i.e. i) and the sample
scale (i.e. ii-v).\\ Typical advantages of the process are:

\begin{itemize}
    \item high purity products;

    \item low equipment, operation and processing
costs;
    \item extremely short processing times;
    \item possibility
 of forming unique metastable phases with improved properties as a result of
 the inherent strong non-equilibrium conditions (i.e. steep thermal gradients
 and high heating/cooling rates).
\end{itemize}

However, beside the above advantages, the rapidity of SHS and the
complexity of involved concomitant physical and chemical
phenomena, makes rather difficult the optimization of the process
as well as that of the chemistry and the morphology of the final
products. In addition, the strong exothermicity of the reaction
may generate a combustion wave which passes through all the pellet
by igniting abnormally the reaction thus stopping or making
unstable the reaction.

It is therefore desirable the development of flexible predictive
simulation tools in order to give a significant step forward to
the progress of the SHS process.\\ In this study we focus our
attention to a specific Ti and C particles mixture system, in
order to obtain TiC particles. According to standard
classification, this powder system involves a solid-solid reaction
in the sense that no gas reactant contributes to the combustion.

Many studies have been carried out on the SHS synthesis of TiC as
model material under both microgravity \cite{Makino,Tanabe} and
normal gravity \cite{Kanury,Lakshmikantha} conditions, either
experimentally \cite{Makino,Tanabe} and theoretically
\cite{Kanury,Lakshmikantha}. This system is particularly
convenient because Ti and C powders are not very expensive. In
addition, the formation heat of TiC is very large (185 kJ/mol) and
the melting point of the product is very high (3423 K) which make
the reactions initiation rather easy and the product formation
straightforward.

The study of this reacting system is complex because of the many
processes involved e.g. the propagation of the combustion wave in
the given granular system, the phase changes of the metallic
component, the occurrence of the chemical reaction, the physical
and mechanical interaction among the powders and the heat and mass
transport in the system which implies strongly inhomogeneous
properties.

We propose a discrete approach to model the kinetics of the
reaction considering the thermal and the chemical evolution of a
small portion of the system, thereby simulating the actual non
uniformity of the mixture properties.

In the section II, we describe the physical phenomena we take into
consideration in the modelling. Section III describes the
numerical methods used to reproduce these physics processes at
macroscopic and microscopic level and their relative coupling.
Section IV presents results on the formation of nanoparticles
during a typical combustion process.

\section{Physical processes and mathematical model}

During the SHS operation, the pellets are rapidly heated. Usually,
a thermal wave precedes the combustion front, thus preheating the
powder mixture ahead. Usually, the metallic constituent has to
melt before reaction occurs \cite{Dunmead}. Under this condition,
the reactions is mainly controlled by the carbon diffusion in the
liquid Ti.

The reaction scheme is

\begin{equation}
Ti(liq)+C(sol)\rightarrow TiC(sol)
\vspace{12pt}\label{react_scheme}\end{equation}

All the atoms and molecules, and in particular the composites of
titanium carbide, are subjects to an interaction potential with
all the other species and at the same time they are thermically
perturbed by the Brownian motion. In consequence of these
competing effects nanoparticles of TiC can be formed.

In the developed model the following steps are considered:

\begin{description}
    \item[a)] the solid-liquid transition of titanium;
    \item[b)] the chemical reaction between the melted titanium and
    the graphitic carbon granules;
    \item[c)] the interaction between clusters of different kinds of
    particles (Ti, C and TiC);
    \item[d)] the formation of nano-sized aggregates of TiC.
\end{description}

Below we analyze in detail  the three fundamental steps: phase
change, chemical reaction, nanoparticle formation.

\subsection{Heat transfer and phase change}

\begin{figure}
\centering
\includegraphics[width=70mm,height=50mm]{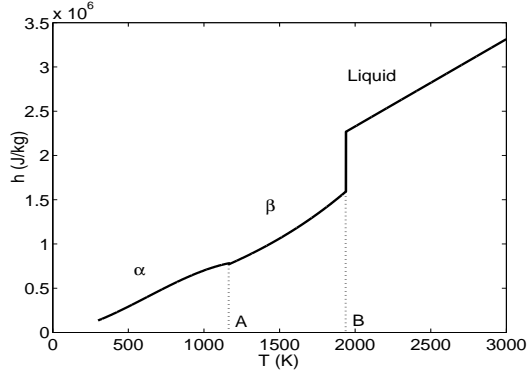}
\caption{Enthalpy per unit of mass versus temperature curve for
titanium}\label{H_T_art}
\end{figure}

In the solid-liquid transition the enthalpy of the system has to
be taken into consideration. In figure \ref{H_T_art} the curve of
the enthalpy per unit of mass ($h$) as a function of the
temperature is shown for titanium.

The function $h(T)$ includes three continuous parts divided by two
discontinuity points (i.e. A and B) at the $\alpha-\beta$ and
$\beta$-liquid transition, which occur at $T_{\alpha,\beta}=1155 $
K and $T_{M,Ti}=1939$ K. During melting or $\alpha-\beta$
transition it is necessary to supply an extra amount of enthalpy
to break the interatomic bonds between the atoms of the solid
metal ($L_{m}=295.555$ kJ/kg and $L_{\alpha,\beta}=59.247$ kJ/kg respectively).\\
The enthalpy versus temperature function for a mass $m$ is defined
as

\begin{equation}
\mathcal{H}(T)=mh(T). \vspace{12pt}
\end{equation}

The heat conduction equation for the  system is given by
\cite{Carslaw}:

\begin{equation}
\rho{C}\frac{\partial T}{\partial t}= \nabla k \nabla T+S,
\vspace{12pt}\label{heat-eq}
\end{equation} where $T$ is temperature, $S$ the energy
source, $\rho$ the density, $C$ the specific heat at constant
pressure and $k$ the thermal conductivity.

However, equation (\ref{heat-eq}) it is not convenient in this
form when $\mathcal{H}(T)$ is discontinuous.  In this condition
the presence of a heat flux does not modify the temperature in the
volume $V$ where phase change is taking place and the eq.
(\ref{heat-eq}) can be replaced more conveniently by:

\begin{equation}
\frac{d}{dt}\int_{V}h\rho=\oint_{\partial V} q
\label{eq-camb-fase} \vspace{12pt}\end{equation} where $q$ is the
specific heat flux (J/$m^{2}$s) across its surface given by the
standard Fourier law

\begin{equation}
q=k\nabla T.\label{fourier} \vspace{12pt}\end{equation}

\subsection{Chemical reaction}

As experiments suggest we assume that the chemical reaction occurs
when the Ti surrounding the C particles has melted, while the
carbon remains solid since its melting point, $T_{M,C}=5000$ K, is
far higher temperature. Similarly the TiC being created
is solid since its melting temperature is $T_{M,TiC}=3423$ K.\\
Due to the exothermicity of the reaction, the heat source in eq.
(\ref{heat-eq}) is evaluated as

\begin{equation}
S=m_{TiC}\Delta H_{r},\label{source}
\vspace{12pt}\end{equation}where $m_{TiC}$ is the mass of the TiC
product and $\Delta H_{r}$ indicates the heat generated per unit
of mass of product($\Delta H_{r}=3.08\cdot10^{3}$ kJ/kg
\cite{por_dil}).

\subsection{Nanoclusters interactions}

After the reaction has taken place, the temperature of the system
is locally increased and the TiC product becomes dispersed in the
liquid Ti phase. Since the atomistic simulation of Ti, C and TiC
is computationally prohibitively expensive, we consider the
dynamics of clusters made of a large number of atoms. The behavior
of the clusters attemps to reproduce on the average that of the
single particles relatively to the phenomena we study.
\\ The motion of the cluster is partly deterministic (due to the
interaction with the other clusters) and partly stochastic,
because of their Brownian motion.\\ For the particles $l$ and $m$
the potential energy has the form \cite{Rapaport}

\begin{equation}
\Psi(r_{lm})=4\epsilon\left[\left(\frac{\sigma}{r_{lm}}\right)^{12}-\left(\frac{\sigma}{r_{lm}}\right)^{6}\right],\quad
r_{lm}\leq r_{c},\label{lenn_jon}\vspace{12pt}
\end{equation}
where $\textbf{r}_{lm}=\textbf{r}_{l}-\textbf{r}_{m}$ and $r=
|\textbf{r}|$. The parameter $\epsilon$ defines the strength of
the pair interaction and $\sigma$ defines a length scale
\cite{Rapaport}. The interaction is repulsive at close distance,
and attractive at larger distance. A cut off is assumed, for
numerical convenience, beyond a limiting separation $r_{c}$. The
corresponding \emph{Lennard-Jones} force is $\textbf{F}=-\nabla
\Psi(r)$ and has the expression
\begin{equation}
\textbf{F}_{lm}=\left(\frac{48\epsilon}{\sigma^{2}}\right)\left[\left(\frac{\sigma}{r_{lm}}\right)^{14}-\frac{1}{2}\left(\frac{\sigma}{r_{lm}}\right)^{8}\right]\textbf{r}_{lm}.
\vspace{12pt}\end{equation}The detailed motion of the clusters is
described by the \emph{Langevin} equation \cite{Reif}

\begin{equation}
m\frac{d\textbf{v}}{dt}=-\alpha \textbf{v}+\textbf{F}^{*}(t)\label{eq-lang}\\
\vspace{12pt}\end{equation} where $m$ is the cluster mass,
$\textbf{v}$ the velocity, $\alpha$ a parameter depending on the
liquid \cite{Reif} and $\textbf{F}^{*}(t)$ the Brownian random
force acting on the cluster.

The motion of these clusters, and in particular that of the TiC
clusters, can generate a phenomenon of coalescence in the form of
very small aggregates of nano-metric sizes. Thus, starting from
micro-sized reacting particles of carbon, we obtain nano-sized
aggregates of TiC.

\section{Numerical methods}

The simulation of the processes described above and the
discretization of the various mathematical models used to describe
them requires to handle multiple length and time scales. The
Brownian motion and the chemical reactions are much slower than
the evolution of the temperature. The processes at the particle
level (chemical reactions, motion) require a spatial resolution at
the nanometer range, while the temperature field is characterized
by micrometer scales and the overall system is on the centimeter
scale.

\begin{figure}
    \centering {\includegraphics[angle=90,width=150mm]{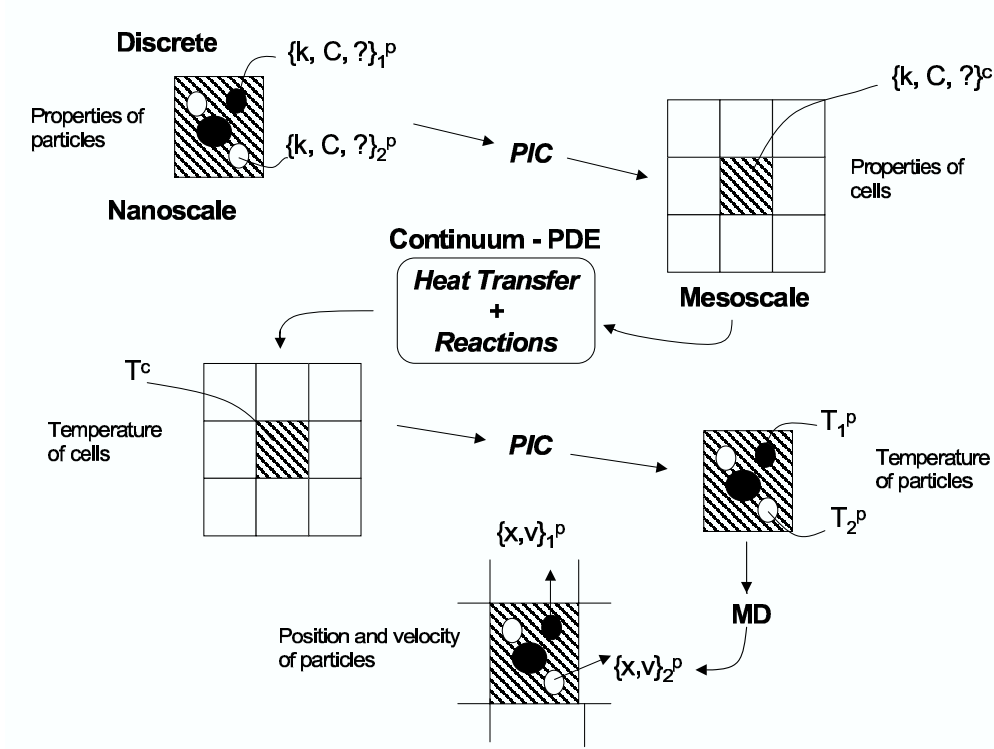}}
    \caption{Coupling of the continuous and the particle models}
\label{sommario}
\end{figure}

To handle these multiple scales we have decided to focus our
attention at the mesoscopic level, describing the evolution of
only a portion of the complete system. Our domain is on the
micrometer range and is treated at that level using a continuum
model discretized with a finite volume (FV) method. However, we
retain the physics at the nanometer range by resorting to two
additional methods: we describe the nanometer physics using
computational particles. The interactions among particles at the
nanometer range are treated with the molecular dynamics (MD)
approach, and the coupling between mesoscopic (micrometer) scales
and nanometer scales is treated with the particle in cell (PIC)
method. Figure~\ref{sommario} summarizes the approach.

This approach has the advantage to use the numerical approach most
suitable for each phenomenon we consider but requires to handle
multiple methods and algorithms and their mutual interaction. To
this end we have applied a modular approach based on the
object-oriented software paradigm.

\subsection{Mesoscopic model}

The heat transfer in a system with  phase changes is described by
equation (\ref{heat-eq}) coupled to equation (\ref{fourier}). It
is a continuum model that involves a local energy balance. At each
point of the system the enthalpy content, which depends on the
local temperature and on the state of aggregation of the matter,
is known.\\ When a chemical reaction event occurs, the amount
$\Delta H_{r}$ of energy released is added to the local enthalpy
content as a source term in eq. (\ref{heat-eq}). Consequently the
local temperature is increased. The TiC particles generated become
a part of the system.

The integration of (\ref{heat-eq}) and (\ref{fourier}) is the
obtained using a \emph{finite volume} method on a two dimensional
cartesian domain $\Omega$ of sizes $l_{x}$, $l_{y}$ with a
boundary indicated by $\Gamma$. This system is discretized using a
two dimensional uniform grid in the direction $x$ and $y$ with
cells of sizes $\Delta x$ and $\Delta y$. We also defined a third
dimension of size $l_{z}$ which is used to dimensionalize all
quantities as in the physical system. The time is discretized in
time steps $\Delta t$.\\ The discretized forms of eq.
(\ref{eq-camb-fase}) and (\ref{heat-eq}) are applied to cells with
or without phase change respectively. To avoid numerical
instabilities we use an implicit scheme based on the Euler
algorithm.

\begin{itemize}
    \item cells with no phase change
\begin{eqnarray}
&C_{ij}^{n}&\rho_{ij}^{n}\frac {T_{ij}^{n+1}-T_{ij}^{n}}{\Delta
t}=k_{ij}^{n}\frac
{T_{i+1,j}^{n+1}+T_{i-1,j}^{n+1}-2T_{ij}^{n+1}}{\Delta x^2}+\label{schema_2d}\\
&+&k_{ij}^{n}\frac{T_{i,j+1}^{n+1}+T_{i,j-1}^{n+1}-2T_{i,j}^{n+1}}{\Delta
y^{2}}+S_{ij}^{n+1} \nonumber\vspace{12pt}
\end{eqnarray}
    \item cells with phase change

\begin{equation}
\Delta H_{ij}^{n}=A_{ij} \Delta t q_{ij}^{n},
\vspace{12pt}\end{equation} being $A_{ij}=2l_{z}(\Delta x + \Delta
y)$ the perimeter of the cell in which phase change is taking
place. The heat flux is calculated by the discretized form of eq.
(\ref{fourier}):

\begin{eqnarray}\label{fourier_discr}
q_{ij}^{n}=k_{i-1,j}^{n*}\frac{T_{ij}^{n}-T_{i-1,j}^{n}}{\Delta
x}+k_{i+1,j}^{n*}\frac{T_{ij}^{n}-T_{i+1,j}^{n}}{\Delta x} \\
+k_{i,j-1}^{n*}\frac{T_{ij}^{n}-T_{i,j-1}^{n}}{\Delta
y}+k_{i,j+1}^{n*}\frac{T_{ij}^{n}-T_{i,j+1}^{n}}{\Delta
y},\nonumber \vspace{12pt}\end{eqnarray} where $k_{i-1,j}^{n*}$ is
the thermal conductivity at the interface between the $i,j$ and
the $i-1,j$ cells. Its value is given by

\begin{equation}\
k_{i-1,j}^{n*}=\frac{k_{ij}^{n}+k_{i-1,j}^{n}}{2}.
\vspace{12pt}\label{cond_discr}\end{equation}Analogous expressions
are valid for all other interfaces.
\end{itemize}

Upon computing the coefficients $k_{ij}$ at $n$ the time level ,
the eqs. (\ref{schema_2d})-(\ref{cond_discr}) become a linear
system which is solved using the \emph{GMRES} algorithm.

A new algorithm has been implemented to solve the problems arising
from the non-linearity and discontinuity of the enthalpy function
versus temperature, $\mathcal{H}(T)$. In order to illustrate the
approach, we consider for convenience solidification first, and
then melting for a given cell of the system.

\subsubsection{Solidification}

For temperatures above $T_{M,Ti}$, we use eq. (\ref{schema_2d}) to
advance the temperature $T_{ij}^{n}$ for the cell $ij$ at the
temporal step $n$ , the corresponding enthalpy content for a
generic cell is given by

\begin{equation}
H_{ij}^{n}=\mathcal{H}(T_{ij}^{n}).
\vspace{12pt}\label{ent_temp_alta}\end{equation} We indicate with
$H_{L}$ and $H_{I}$ ($H_{L}>H_{I}$) the enthalpies of a cell of
titanium corresponding to the beginning and the end of the
solidification respectively.

Solidification starts when for $n=\widetilde{n}$ we have
\begin{equation}
\begin{array}{l}
H_{ij}^{\widetilde{n}}>H_{L}\\
T_{ij}^{\widetilde{n}+1}<T_{M,Ti}\\
H_{ij}^{\widetilde{n}+1}<H_{I} \vspace{12pt}\end{array}
\end{equation}and at subsequent time levels $m\geq \widetilde{n}+1$
, the following condition holds:
\begin{equation}
\begin{array}{l}
H_{ij}^{m}\leq H_{L}\\
H_{ij}^{m}\geq H_{I},
\vspace{12pt}\label{cond_4}\end{array}\end{equation}

In that case, eq. (\ref{schema_2d}) is no longer appropriate and
the enthalpy and the temperature are advanced using:
\begin{equation}
\begin{cases}
H_{ij}^{m}=H_{ij}^{\widetilde{n}}+ A_{ij}\Delta t \sum_{k=\widetilde{n}+1}^{m}q_{ij}^{k} \\
T_{ij}^{m}=T_{M,Ti}.
\end{cases} \vspace{12pt}\label{ent_temp_pass} \end{equation}

Finally, the solidification process is completed at the time step
$m=\widetilde{m}$, when the following condition holds:
\begin{equation}
H_{ij}^{\widetilde{m}+1}<H_{I}<H_{ij}^{\widetilde{m}}
\vspace{12pt}\end{equation}

In this last instance, the temperature is advanced as

\begin{equation}
T_{ij}^{\widetilde{m}+1}=\mathcal{H}^{-1}(T_{ij}^{\widetilde{m}+1}).
\vspace{12pt}\end{equation}

Later, solidification no longer occurs in the cell and, for time
steps $r>\widetilde{m}+1$, the temperature is again computed using
eq. (\ref{schema_2d}) and the enthalpy is given by
\begin{equation}
H_{ij}^{r}=\mathcal{H}(T_{ij}^{r}).\label{ent_T_bassa}
\vspace{12pt}\end{equation}

\subsubsection{Melting}

In the case of melting, the system starts at a temperature below
$T_{M,Ti}$ and eq. (\ref{schema_2d}) governs the evolution of the
temperature. Again the enthalpy is computed from
\begin{equation}
H_{ij}^{u}=\mathcal{H}(T_{ij}^{u}) \vspace{12pt}\end{equation}

The process of solidification starts at the time step
$u=\widetilde{u}$, when the following conditions are verified
\begin{equation}
\begin{array}{l}
H_{ij}^{\widetilde{u}}<H_{I}\\
T_{ij}^{\widetilde{u}+1}>T_{M,Ti}\\
H_{ij}^{\widetilde{u}+1}>H_{L}
\end{array}
\vspace{12pt}\end{equation}

Subsequently, for $v\geq \widetilde{u}+1$, enthalpy and
temperature are calculated using
\begin{equation}
\begin{cases}
H_{ij}^{v}=H_{ij}^{\widetilde{u}}+A_{ij}\Delta t \sum_{k=\widetilde{u}+1}^{v}q_{ij}^{k} \\
T_{ij}^{v}=T_{M,Ti}.
\end{cases} \vspace{12pt}\end{equation}
Solidification ends at the time step $v=\widetilde{v}$, when the
condition below is verified
\begin{equation}
H_{ij}^{\widetilde{v}}<H_{L}<H_{ij}^{\widetilde{v}+1}
\vspace{12pt}\end{equation} At the time step $\widetilde{v}+1$,
the temperature is advanced as prescribed by conservation of
energy:
\begin{equation}
T_{ij}^{\widetilde{v}+1}=\mathcal{H}^{-1}(T_{ij}^{\widetilde{v}+1}).
\vspace{12pt}\end{equation}

Finally when melting is completed, for $z>\widetilde{v}+1$ the
temperature is advanced using eq. (\ref{schema_2d}) and the
enthalpy is given by

\begin{equation}
H_{ij}^{z}=\mathcal{H}(T_{ij}^{z}) \vspace{12pt}\end{equation}

\subsection{Coupling of mesoscopic and
nanometer scale through PIC}

The mesoscopic subdivision of the system is made using the grid of
cells described above. The cells are used to treat the heat
transfer in a melting or solidifying continuum media.\\ To study
the nanometer scale physics and the detail of the phenomena at
particle (i.e. nanometer) level, it is necessary to use a discrete
approach. Thus, the \emph{Particle In Cell} method (PIC)
\cite{Birdsall} has been adopted. This implies that the system is
subdivided in computational particles which physically represent a
cluster of atoms. Therefore the granules of carbon correspond to a
set of computational particles with the properties of the carbon
atoms. Similarly the Ti and the TiC are represented by
computational particles with the corresponding properties. Then
the cell properties are computed based on the particles properties
which change in time. This approach is particularly convenient in
the study of powder systems when the mass transfer and the heat
transport are closely coupled.

The two levels of description of the system permit us to map not
only the thermophysical properties ($C$, $\rho$ and $k$) from one
level to the other, but also to map the field variables such as
temperature and enthalpy. Thus the temperature and the enthalpy of
a particle can be computed by knowing the corresponding values in
the cells and viceversa.

The approach followed is summarized in Fig.~\ref{sommario}. The
system is characterized by $N_{p}$ particles, each of volume
$V_{p}$, and concurrently the same system can also be represented
by cells of volume $V_{c}$.\\ Assuming an equal number of
particles per cell, $N_{pc}$, the initial volume of each particle
is given by

\begin{equation}
V_{p}=\frac{V_{c}}{N_{pc}}. \vspace{12pt}\end{equation} By
introducing a system of logical coordinates as
$\xi=\frac{x}{\Delta x}$ and $\eta=\frac{y}{\Delta y}$, for a
generic cell $c$ and particle $p$ we define the \emph{assignment
function W} as the function resulting from the tensor product
between $b-spline$ functions of first order in each direction:

\begin{equation}
W(\xi_{c}-\xi_{p},\eta_{c}-\eta_{p})=b_{1}(\xi_{c}-\xi_{p})b_{1}(\eta_{c}-\eta_{p}),
\vspace{12pt}\end{equation} where ($\xi_{p},\eta_{p}$) is the
position of the particle, ($\xi_{c},\eta_{c}$) the position of the
cell center and $b_{1}$ is given by
\begin{equation}
b_{1}(\xi)=
\begin{cases}
1-|\xi| \quad {\rm if} \quad |\xi|<1\\
0 \quad {\rm otherwise}.
\end{cases}
\vspace{12pt}\end{equation} This function which will be denoted by
$W_{cp}$ for shorthand, gives the contribution of a generic
property of the particle $p$ to the cell $c$ as a function of the
relative positions.
\\ Similarly, it is possible to perform the opposite operation
consisting in mapping a property (or a physical quantity) of the
cells to a particle.

The properties and the physical quantities that are calculated
with this method are summarized in Table I.
\begin{table}
\centering
\begin{tabular}{|c|c|}

  \hline
  From particles to cells & From cells to particles \\
  \hline
  $\displaystyle\rho_{c}=\frac{\sum_{p}\rho_{p}V_{p}W_{cp}}{\sum_{p}V_{p}W_{cp}}$ & $\displaystyle T_{p}=\sum_{c}W_{cp}T_{c}$ \\
  \hline
  $\displaystyle K_{c}=\frac{\sum_{p}K_{p}V_{p}W_{cp}}{\sum_{p}V_{p}W_{cp}}$ &  \\
  \hline
  $\displaystyle C_{p,c}=\frac{\sum_{p}C_{p,p}V_{p}W_{cp}}{\sum_{p}V_{p}W_{cp}}$ &  \\
  \hline
\end{tabular}
\caption{Properties and physical quantities calculated using PIC}
\end{table}
By knowing the temperature in each cell, at each time step it is
possible to single out those where the reaction can take place. By
recalling the basic assumption, e.g. the titanium must be melted
before the reaction can take place, the condition that must be
verified is

\begin{equation}
T_{ij}^{n}>T_{M,Ti}\label{cond_react}. \vspace{12pt}\end{equation}
Another necessary condition in order to have a reaction in the
cells which fulfil the condition (\ref{cond_react}), is that each
cell contains at least one particle of titanium and one of carbon.
If more than one couple is present, then the reaction occurs first
between those particles which have the shortest distance. These
two particles are thus replaced by two new ones. If we indicate
with ${Ti}$ the particle of titanium and with ${C}$ that of
carbon, the number of moles of the new TiC particle ${TiC}$ being
created is

\begin{equation}
\begin{cases}
n_{TiC}=n_{Ti}\quad {\rm if}\quad n_{Ti}<n_{C}\\
n_{TiC}=n_{C}\quad  {\rm if}\quad n_{Ti}>n_{C}\\
n_{TiC}=n_{Ti}\quad {\rm if}\quad n_{Ti}=n_{C}
\end{cases}
\vspace{12pt}\end{equation}whereas its volume is

\begin{equation}
V_{TiC}=\frac{n_{TiC}(PM_{Ti}+PM_{C})}{\rho_{TiC}},
\vspace{12pt}\end{equation}where $PM_{Ti}$ and $PM_{C}$ are the
molecular weights of titanium and carbon.\\ For sake of mass
balance a new particle must be introduced into the system. This
can be either C or Ti particle. Its volume must fulfils the
conditions (\ref{cond_mass})

\begin{equation}
\begin{cases}
V_{C}=\frac{(n_{C}-n_{TiC})PM_{C}}{\rho_{C}}\quad {\rm if}\quad n_{Ti}<n_{C}\\
V_{Ti}=\frac{(n_{Ti}-n_{TiC})PM_{Ti}}{\rho_{Ti}}\quad {\rm
if}\quad n_{Ti}>n_{C}
\end{cases}\label{cond_mass}
\vspace{12pt}
\end{equation}
If the condition $n_{Ti}=n_{C}$  is verified, only a particle of
TiC is created.
\\ The amount of energy developed during the formation of the
TiC particle is calculated using eq.~(\ref{source}).

\subsection{Molecular Dynamics treatment of nanometer scale particles}

The interaction between different particles in the system and
among the atoms that compose them is an aspect which plays a very
important role in the phenomenon of the formation of
nanoparticles. This interaction is reproduced by the Lennard-Jones
potential. This is only a first approximation of the real
situation, because it usually holds for atoms or molecules and not
for clusters. Indeed, our computational particles represent
cluster of atoms that, on average, we assume interacting with a
potential that includes a repulsive and an attractive part as that
of Lennard-Jones. In calculating the potential we consider the
distance between the center of the clusters.\\ The equations of
the motion for a generic computational particle are

\begin{eqnarray}
m\frac{d\textbf{v}}{dt}&=&\textbf{F}(t)\\
\frac{d\textbf{x}}{dt}&=&\textbf{v},\label{eq-moto}
\vspace{12pt}\end{eqnarray} where $\textbf{x}$, $\textbf{v}$ and
\textbf{F} are position, velocity and force for a particle of mass
$m$. The equations of motion are solved using the Verlet algorithm
\cite{Frenkel}. To simulate the stochastic behaviour due to the
Brownian motion of the particle we use the \emph{Andersen
thermostat} approach \cite{Frenkel}. We consider each cell as
interacting with an \emph{heat bath} which sets its temperature at
the value given by the continuum model described above while, at
the same time, it permits this subsystem to access all the energy
shells corresponding to this temperature according to their
Boltzmann weight \cite{Frenkel}. The coupling of the system to the
bath is represented by stochastic velocities which are
occasionally randomly assigned to the computational particles,
using a collision probability. The strength of this coupling
depends on the frequency $\nu$ of stochastic collisions between
particles. The values of the accessible velocities are those of
the Maxwell-Boltzmann distribution \cite{Reif} with variance

\begin{equation}
\sigma=\sqrt{\frac{k_{B}T_{p}}{m}},
\vspace{12pt}\end{equation}where $k_{B}$ is the Boltzmann constant
and $T_{p}$ is the temperature of the particle.

\section{Results of the simulations}

\subsection{Validation test - 1D analytical benchmark}

At first, we performed a comparison between the results of the
present model with an analytical solution for a system undergoing
a phase change~\cite{Carslaw}. The simulation has been carried out
on a one-dimensional domain of length $l=0.1~m$ made of titanium
for a time $t_{f}=150~s$ with the following initial and boundary
conditions

\begin{equation}
\begin{cases}
T_{i}^{0}=0\quad\forall i\\
T_{0}^{n}=2500\quad\forall n\\
T_{l}^{n}=0\quad\forall n
\end{cases}
\vspace{12pt}\end{equation}

\begin{figure}
\centering
\includegraphics[width=70mm,height=50mm]{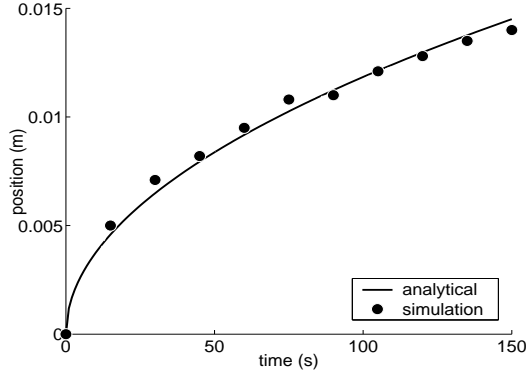}
\caption{Comparison between model prediction (dashed line) and
analytical solution (solid line)}\label{test_libro}
\end{figure}

In fig.~\ref{test_libro} we show the time evolution of the
liquid-solid interface. As it is possibile to see the agreement
with the analytical solution is good.
 We have conducted a convergence study to verify the correctness of the implementation of the heat transport algorithm described above.

\subsection{Validation test - 2D phase change benchmark}

We carried out a simulation for a two-dimensional system made of
titanium with sizes $l_{x}=l_{y}=0.015~m$ for a time $t_{f}=5~s$.
The initial and boundary conditions were

\begin{equation}
\begin{cases}
T_{i}^{0}=300+3000sin\left(\frac{\pi}{l_{x}}x_{i}\right)\quad\forall i\\
T_{\Gamma}^{n}=300\quad\forall n.\end{cases}
\vspace{12pt}\end{equation}

\begin{figure}
\centering
\includegraphics[width=70mm,height=50mm]{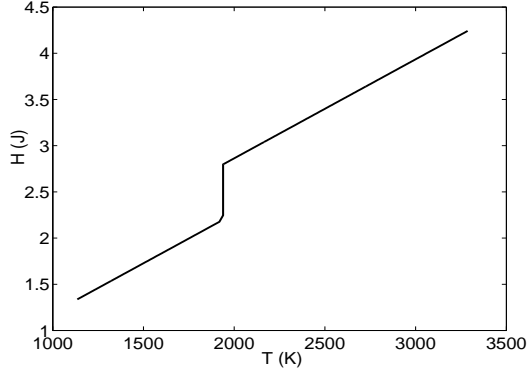}
\caption{Computed enthalpy for the central cell of a solidifying
system made of titanium}\label{ent_cell}
\end{figure}

The enthalpy versus the temperature for the central cell is shown
in fig.~\ref{ent_cell}. The result proves that the phase change
happens correctly and with the correct enthalpy change
corresponding to the latent heat.

\subsection{Formation kinetics of TiC nanoparticles}

In this case the system considered is composed by titanium
surrounding three granules of graphite,  two with a diameter
$1.5~\mu m$ and the other with diameter $1~\mu m$. The size
of each side of the domain is $l_{x}=l_{y}=10~\mu m$.\\
The properties of the materials, and in particular the specific
heat and the thermal conductivity, depend on the temperature and
are obtained from \cite{Toluokian1,Toluokian2}. The density is
assumed as constant for both materials and equal to that at
$300~K$. This means that the effects of the volume variation of
the species during the process are neglected.

A generic heat source is introduced to simulate the melting of
titanium. The thermal behavior of the system assumes the following
initial and boundary conditions

\begin{equation}
\begin{cases}
T_{ij}^{0}=2600+500 \sin\left(\frac{\pi}{l_{x}}x_{ij}\right)\quad\forall i,\forall j\\
T_{\Gamma}^{n}=2600\quad\forall n.\end{cases}
\vspace{12pt}\end{equation}

The domain is subdivided by using $N_{x}$ and $N_{y}$ cells along
$x$ and $y$ directions respectively and with $N_{pc}$ particles
per cell. The simulation time is $t_{f}$, the numbers of time
steps in the heat equation and in the motion equations are
$n_{th}$ and $n_{m}$ respectively. Subcycling of the heat equation
is used to handle its much faster scale, compared with the scale
of the particle motion. For the simulation shown below, the
following values were used: $N_{x}=10$, $N_{y}=10$, $N_{pc}=9$,
$t_{f}=1\cdot10^{-1}~s$, $n_{th}=1500$ and $n_{m}=50000$. The
initial velocity of all particles is zero, allowing the Andersen
thermostat to establish the proper thermal equilibrium. The
collision frequency used in the Andersen thermostat is
$\nu=25~kHz$.

\begin{figure}
\begin{tabular}{cc}
  \includegraphics[width=70mm,height=70mm]{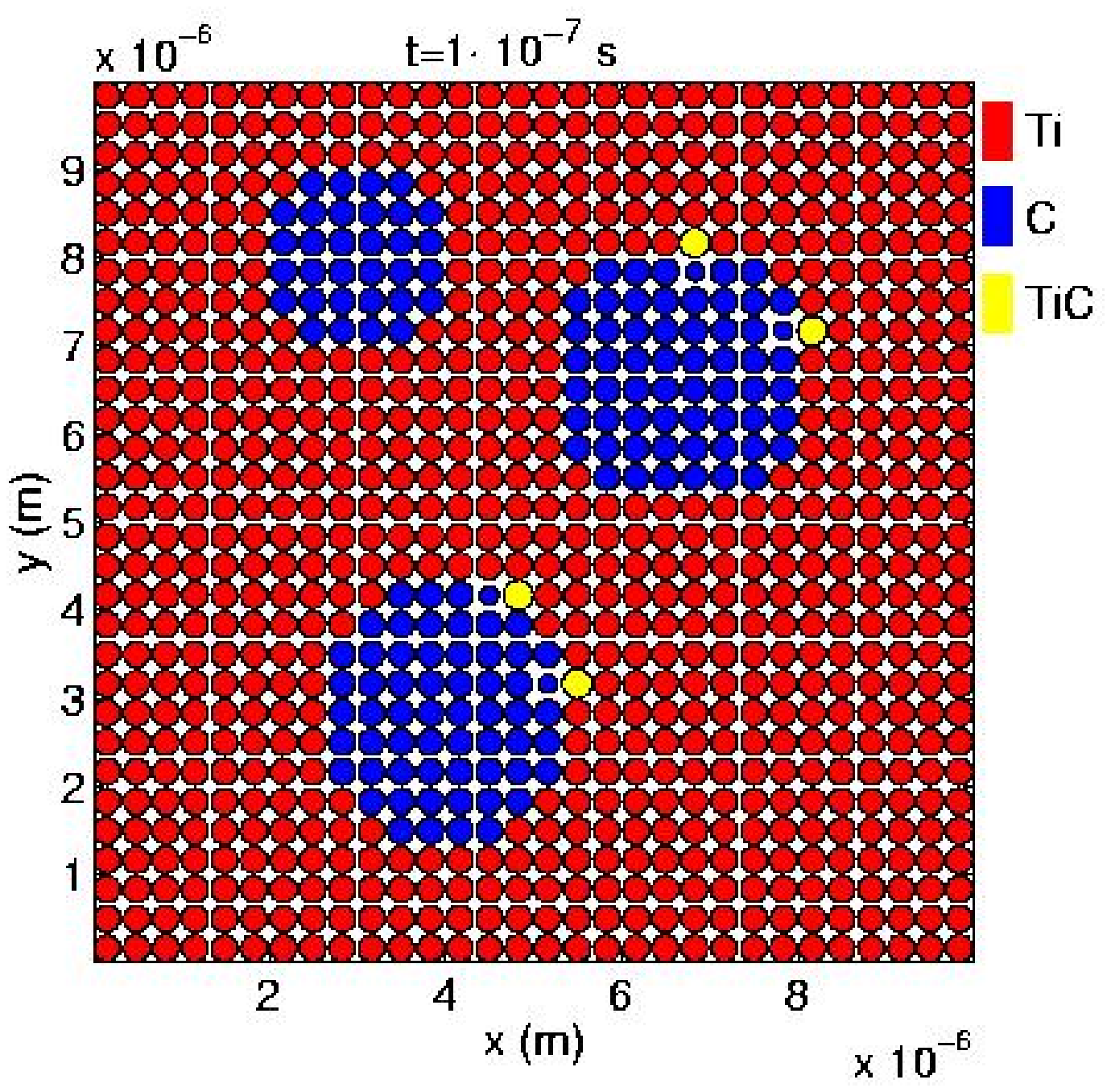}

  &    \includegraphics[width=70mm,height=70mm]{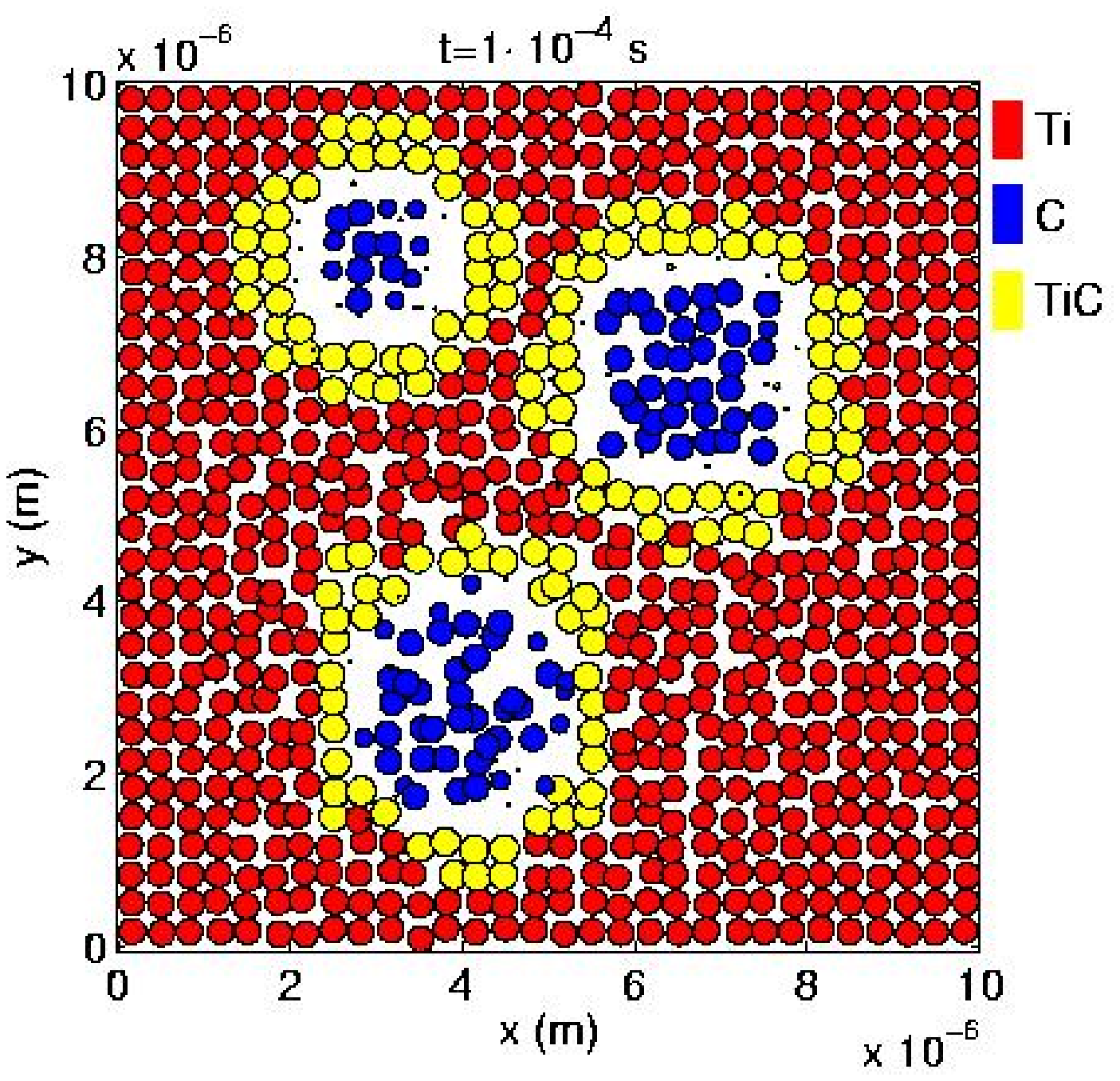} \\
     \includegraphics[width=70mm,height=70mm]{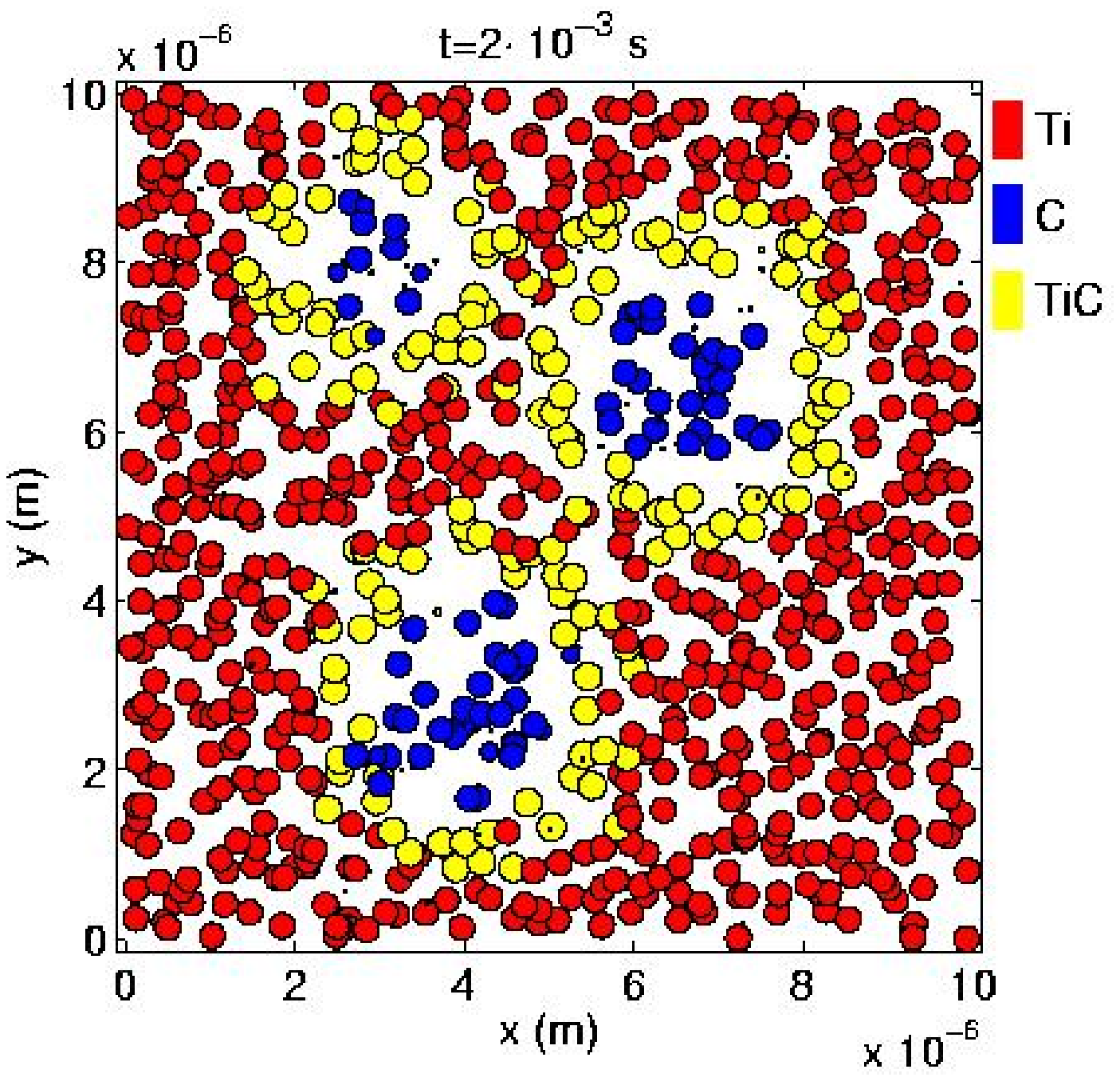} &
    \includegraphics[width=70mm,height=70mm]{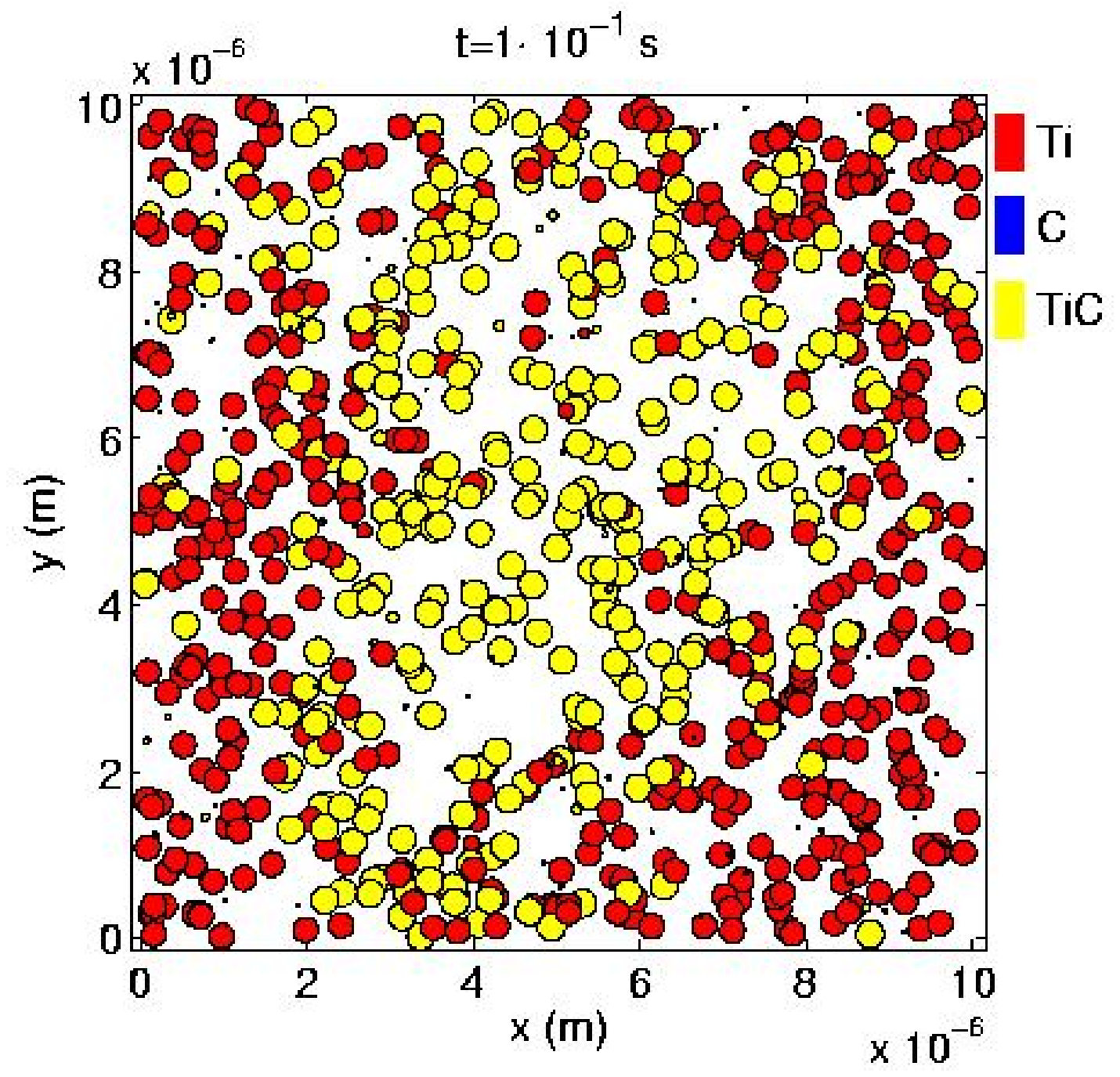}
  \end{tabular}
    \caption{Evolution of the system}\label{evolution}
  \end{figure}

The time evolution of the reacting system is shown in
fig.~\ref{evolution}. The evolution of the combustion process is
evident: TiC shells of particles are first formed at the interface
between graphite and liquid Ti, as demonstrated in the
experiments. After this stage, the thermal agitation of the
particles breaks the shells and permits that the reaction develops
further. The final aggregates of particles are the end product of
the reaction: TiC nanoparticles.

\begin{figure}
    \centering {\includegraphics[width=70mm,height=50mm]{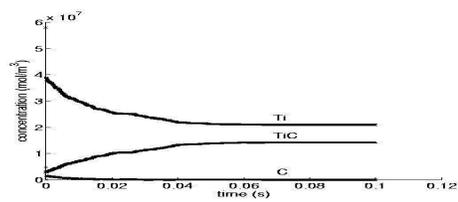}}
    \caption{Evolution of the concentration of the chemical
    species}\label{concentration}
\end{figure}

Figure~\ref{concentration} shows the evolution of the
concentration of the various chemical species for the same
simulation.

\section{Conclusions}

A new coupled mesoscopic-nanometer scale model has been developed
to simulate the SHS process. The approach combines the solution of
PDEs for the continuum media on the mesoscopic level with the
discrete PIC-MD techniques for the nanometer scale. This model
permits the description of thermal phenomena on the mesoscale and
physical and chemical interaction on the nanoscale of both
reagents and products species.\\ The overall model offers a unique
opportunity to follow the evolution of the nanostructure during
the SHS process.\\ An interesting output of the meso/nano model is
the concentration history of chemical species which are relatively
easy accessible by actual experiments.\\ The model has been
validated against an analytical solution. The PIC-MD model
requires direct experimentation in order to determine the most
suitable \emph{LJ} parameters and the collision frequency value
for more realistic application of the developed model.

\section*{Acknowledgments}

This research is supported by the United States Department of
Energy, under contract W-7405-ENG-36.


\begin{thebibliography}{99}

\bibitem{Frenkel} D. Frenkel, B. Smit, {\it Understanding Molecular
Simulation} Academic Press, San Diego (1996).

\bibitem{Birdsall}  R.W. Hockney, J.W. Eastwood, {\it
 Computer simulation using particles} A. Hilger, Bristol (1988).

\bibitem{Rapaport} D.C. Rapaport, {\it The Art of Molecular Dynamics Simulation}
Cambridge Univ. Press, Cambridge (1995).

\bibitem{Toluokian1} Y.S. Touloukian, D.P. DeWitt, {\it Thermal Radiative Properties: Metallic Elements and Alloys} IFI/Plenum, New York (1970).

\bibitem{Toluokian2} Y.S. Touloukian, D.P. DeWitt, {\it Thermal Radiative Properties: Nonmetallic Solids} IFI/Plenum, New York (1970).

\bibitem{Comb_proc} A.W. Weimer, {\it Carbide, Nitride and Boride Synthesis and Processing}
Chapman and Hall, London (1997).

\bibitem{por_dil} M.G. Lakshmikantha, J.A. Sekhar, {\it Metallurgical Trans.

A} {\bf 24}, 617 (1993).

\bibitem{Reif} F. Reif, {\it Fundamentals of Statistical and Thermal
Physics} McGraw-Hill, New York (1965).

\bibitem{Carslaw} H.S. Carslaw, J.C. Jaeger, {\it Conduction of Heat in Solids } Oxford University Press, Oxford (1986).

\bibitem{Kanury} A.M. Kanury, {\it Metallurgical Trans. A} {\bf 23}, 2349 (1992).

\bibitem{Merzhanov} A.G. Merzhanov, {\it Combust. Sci. and Tech.} {\bf
98}, 307-336 (1994).

\bibitem{Merzhanov_2} A.G. Merzhanov and I.P. Boroviskaya, {\it Comb. Sci. and Tech. A}
{\bf 10}, 175 (1975).

\bibitem{Suryanarayama} S.B. Badhuri and S. Badhuri, {\it Combustion Synthesis in
Non-equilibrium Processing of Materials} C. Suryanarayama Ed.,
Pergamon Materials Series (1999).

\bibitem{Weiner} A.W. Weiner, {\it Carbide, Nitride and Boride Materials Synthesis and
Processing}, Chapman and Hall (1997).

\bibitem{Dunmead} S.D. Dunmead, D.W. Readey and C.E. Semler, {\it J. Am. Cer.
Soc.} {\bf 72(12)}, 2318 (1989).

\bibitem{Halvenson} D.C. Halvenson, K.H. Ewald and Z. Munir, {\it J. Mat. Sci.} {\bf 28}, 4583
(1993).

\bibitem{Makino} A. Makino, N. Araki, and T. Kuwabara, {\it Trans. Jpn. Soc. Mech.
Eng.} {\bf B58(55)}, 271 (1992).

\bibitem{Lakshmikantha} M.G. Lakshmikantha, A. Bhattacharya and J.A. Sekkar, {\it Met.
Trans.} {\bf A23}, 23 (1992).

\bibitem{Tanabe} Y. Tanabe, T. Sakamoto, N. Okada, T. Akatsu and E. Yasuda, {\it J. Mat. Res.} {\bf 14}, 1516 (1999).





\end{thebibliography}
\end{document}